\documentclass[floats,preprint,aps,amssymb]{revtex4}
\usepackage{graphicx}

\begin{document}

\title{Quantum Extension of the Jarzynski Relation; Analogy with Stochastic
Dephasing}

\author{Shaul Mukamel}

\affiliation{Department of Chemistry and Department of Physics and Astronomy,\\
University of Rochester P. O. RC Box 270216 Rochester, NY
14627-0216}

\vspace{0.5cm}

\date{\today}

\begin{abstract}
\vspace*{1cm} The relation between the distribution of work
performed on a classical system by an external force switched on
an arbitrary timescale, and the corresponding equilibrium free
energy difference, is generalized to quantum systems. Using the
adiabatic representation, we show that this relation holds for
isolated systems as well as for systems coupled to a bath
described by a master equation. A close formal analogy is
established between the present ``classical trajectory'' picture
over populations of adiabatic states and phase fluctuations
(dephasing) of a quantum coherence in spectral lineshapes,
described by the stochastic Liouville equation.

\vspace{2cm} \noindent
 Submitted to \emph{Phys.
Rev. Lett.}

\end{abstract}

\maketitle

%
\section{Introduction}
Jarzynski~\cite{1,3} had established a remarkably simple and
general relationship between the distribution of work performed on
a classical system by an external force and the free energy
difference between the initial and final states~\cite{2,4,6}. This
interesting prediction, connecting equilibrium quantities (free
energies) to nonequilibrium trajectories corresponding to
measurements performed on an arbitrary timescale, has recently
been verified in single molecule measurements~\cite{5}.

In this letter we utilize the adiabatic representation~\cite{9} to
prove that this relation holds equally for quantum systems. For an
isolated driven quantum system (i.e., not coupled to a bath) the
work does not depend on the path, and the proof follows directly
from a sum rule of nonadiabatic couplings. We further develop a
path integral representation of systems coupled to a bath
described by a master equation in the adiabatic  basis using
``trajectories'' over adiabatic state populations, and generalize
Jarzynski's relation to obtain the joint distribution of work and
energy in a nonequillibrium measurement. A close resemblance is
established between the physical picture underlying the
distribution of work and the distribution of phase acquired by a
quantum coherence due to coupling to a bath.

Consider a quantum system coupled to an external force which
changes its Hamiltonian $H(t)$ by switching a parameter $\lambda
(t)$ from $\lambda=0$ at $t=0$ to $\lambda=1$ at time $t$.
Initially the system is in thermal equilibrium and the
instantaneous energy levels (eigenvalues of $H(t)$) are denoted
$\epsilon_n (t)$ with the corresponding eigenstates
$|\varphi_{n}(t)\rangle$. The canonical partition function for a
given value of $\lambda$ is
\begin{equation}
\label{1}
 Z_\lambda = \sum_{n} exp\,[-\beta\epsilon_n(t)]\equiv exp
 (-\beta F_\lambda)
\end{equation}
where $F_\lambda$ is the Helmholtz free energy.

Let us consider an externally driven, but otherwise isolated,
system. In the adiabatic representation \cite{9} we expand its
density matrix as
\begin{equation}
\label{2}
\rho(t)=\sum_{nm}\rho_{nm}(t)|\varphi_n(t)><\varphi_m(t)|,
\end{equation}
where the coefficients $\rho_{nm} (t)$ satisfy the Liouville
equation
\begin{equation}
\label{3}
 \frac{d}{dt}\rho_{kl}(t)= -i \omega_{kl} (t)
\rho_{kl}(t)-\sum_{mn}S_{kl,mn}(t) \rho_{mn}(t)
\end{equation}
with $\omega_{kl}(t) \equiv \epsilon_k (t)-\epsilon_l (t)$ and
\begin{equation}
\label{4}
 S_{kl,mn} (t) =  \langle \varphi_{k}(t)
|\dot{\varphi}_{m}(t) \rangle \delta_{ln}+ \langle
\dot{\varphi}_{n}(t)|\varphi_{l}(t)\rangle\delta_{km}
\end{equation}
The nonadiabatic coupling $S$ is of purely quantum origin and has
no classical analogue.

For a given realization of $\lambda(t)$, the solution of
Eq.~(\ref{3}) allows us to compute the conditional probability
$K_{mn} (t)$ of the system to be in the state $|\varphi_m
(t)\rangle$ at time $t$ given that it started at state $|\varphi_n
(0)\rangle$, so that $\rho_{mm}(t) = K_{mn} (t) \rho_{nn} (0)$.
This probability is normalized by summing over \emph{final} states
$\sum_{m} K_{mn} (t) = 1$. A second, less obvious, sum rule is
obtained by summation over \emph{initial} states $\sum_{n} K_{mn}
(t) = 1$. When the external force varies slowly, it follows from
the adiabatic theorem~\cite{9} that no nonadiabatic transitions
take place, $K_{mn}(t) = \delta_{mn}$, and this sum rule trivially
holds. For finite switching timescales there are nonadiabatic
transitions and the instantaneous eigenstates $|\varphi_n
(t)\rangle$ are no longer the solutions of the time dependent
Schr\"{o}dinger equation. Nevertheless, the second sum rule
follows from the relation
\begin{equation}
\label{second} \sum_{n} S_{kl,nn} (t) =
\frac{d}{dt}\langle\varphi_k (t) | \varphi_l (t) \rangle = 0.
\end{equation}

A direct consequence of Eq.~(\ref{second}) is that the following
uniform distribution in adiabatic population space, obtained by
setting $\rho_{nm}(t) =\delta_{nm}$ at all times in Eq.~(\ref{2})
\begin{equation}
\label{6}
 \rho (t)= \sum_{n}|\varphi_n (t) \rangle\langle\varphi_n
(t)|,
\end{equation}
satisfies the Liouville equation. This interesting identity gives
rise to the second sum rule, which holds despite the nonadiabatic
transitions and is the key for proving the quantum Jarzynski
identity. Alternatively, Eq.~(\ref{6}) can be viewed as a unitary
transformation of the unit operator between adiabatic basis sets
at different times. This equation then simply states that the unit
operator is invariant to a unitary transformation.

We shall now compute the ensemble average of $exp\,(-\beta W)$
where $W$ is the cumulative work performed up to time $t$. If the
system starts in state $|\varphi_n(0)\rangle$ and ends up at time
$t$ in the state $|\varphi_m(t)\rangle$, then the work made by the
external force is $W = \epsilon_m (t)- \epsilon_n (0)$, and we
have
\begin{equation}
\label{7} \langle exp \,(-\beta W)\rangle = \frac{1}{Z_{0}}
\sum_{mn} exp\, [-\beta \epsilon_n (0)] K_{mn} (t) exp\, [-\beta (
\epsilon_{m} (t)-\epsilon_n (0))].
\end{equation}
This gives
\begin{equation}
\label{8}
 \langle exp \,\,(-\beta W)\rangle=\frac{1}{Z_0}\sum _{mn}
K_{mn} (t) exp\,\,[-\beta \epsilon_m (t)].
\end{equation}
Using the second sum rule we can carry out the $n$ summation and
obtain Jarzynski relation,
\begin{equation}
\label{9a} \langle exp \,(-\beta W)\rangle = Z_{1}/Z_{0}
\end{equation}
where $Z_1$ is the final state partition function for $\lambda =
1$. Eq.~(\ref{8}) implies that it is generally possible to compute
the free energy change $F_{1}-F_{0} \equiv -\beta^{-1} \log
(Z_{1}/Z_{0})$ using the distribution of work.

We next turn to a system coupled to a bath and described by a
master equation in the adiabatic basis~\cite{3,6}. We denote the
transition rate from state $n^{\prime}$ to $n$ by
$R_{nn^{\prime}}(t)$. $R$ depends on $t$ since it is recast in the
time dependent, adiabatic, basis. It further satisfies the
detailed balance condition~\cite{13}
\begin{equation}
\label{9}
 R_{nn'}(t)/ R_{n'n}(t) = exp\,\,(-\beta \omega_{nn'}(t))
\end{equation}
For an isolated system the work equals the energy change and is
therefore a state function. When the system is not isolated, the
work becomes path dependent and may not be computed from the
initial and final states alone, as was done in Eq.~(\ref{7}). For
our model, the change in energy along the path stems from the
variation of the eigenvalues $\epsilon_{n}(t)$ which is induced by
the driving force, as well as the changes in their occupations,
which are induced by the coupling to the bath (Fig. 1). The former
is the work whereas the latter is the heat. Both work and heat are
path dependent whereas the overall energy change $\epsilon_m (t) -
\epsilon_n (0)$ is a state function.

We shall compute the joint distribution of work performed $(W)$
and the change in system's energy $(E)$ in the process by defining
the joint energy-work generating function
\begin{equation}
\label{10}
 S (\gamma,\delta;t)= \langle exp \,\,[-\gamma W
(t)-\delta E (t)]\rangle,
\end{equation}
where $\langle....\rangle$ denotes an ensemble average over
trajectories for a fixed realization of $\lambda (t)$. The moments
of $W$ and $E$ can then be computed as derivatives with respect to
the parameters $\gamma$ and $\delta$
\begin{equation}
\label {11}
 \langle W (t)^{p} E^{q} (t) \rangle =
(-1)^{p+q}\frac{\partial^{p}}{\partial
\gamma^{p}}\frac{\partial^{q}}{\partial\delta^{q}} S
(\gamma,\delta;t)\,|_{\gamma= \delta = 0}
\end{equation}

By expanding the solution of the master equation perturbatively in
the off diagonal elements of $R_{nn{^\prime}} (n \neq n^{\prime})$
it can be represented as a path integral in the adiabatic
population space (see Fig. 1): Consider a family of trajectories
which start at state $|\varphi_{n}(0)\rangle$  at $t=0$ and end in
state $|\varphi_{m}(t)\rangle$ at time $t$. In the course of a
given trajectory the system had jumped $j$ times between different
adiabatic states with $j=0,1,2....$. The conditional probability
of the system to start at state $|\varphi_{n}(0) \rangle$ and end
in the state $|\varphi_{m}(t) \rangle$ is then given by a sum over
all possible values of $j$. The work can be computed by
integrating $\partial \epsilon_n/\partial t$ along the continuous
segments of each trajectory. The contribution of this family to
the generating function can be calculated by solving the equation
of motion for the following Green function
\begin{equation}
\label{12} \frac{dG_{mn}(t)}{dt}= \sum_{n^{\prime}}
R_{mn^{\prime}} (t) G_{n\prime n}(t) -
\gamma\frac{\partial\epsilon_m}{\partial t}G_{mn}(t)
\end{equation}
with the initial condition $G_{mn}(0)=\delta_{mn}$. It immediately
follows from the path integral representation of the solution of
the master equation that the generating function is given by
averaging the Green function over the equilibrium distribution of
initial states and a weighted summation over final states
\begin{equation}
\label{13} S (\gamma,\delta;t) = \sum_{nm}exp
[-\delta[\epsilon_{m} (t) - \epsilon_{n} (0)]]  G_{mn} (t;\gamma)
\frac{exp (-\beta\epsilon_{n}(0))}{Z_{0}}.
\end{equation}

Eq.~(\ref{13}) provides the following physical picture for the
process: We consider an ensemble of trajectories over adiabatic
state populations. In the classical case~\cite{1} the ensemble is
generated by the distribution of initial states, whereas here the
ensemble originates from the random nature of the stochastic jumps
among adiabatic states. The Green function represents an open
system where population can flow to (or from) state $m$ with a
rate $-\gamma
\partial\epsilon_{m}/\partial t$ which depends on the variation
of the adiabatic energies with time and on the parameter $\gamma$.
As a result, the total population $\sum_m G_{mn} (t)$ is not
conserved and its value determines the generating function for the
work and energy. The adiabatic representation thus provides a
``classical trajectory'' picture for the joint distribution of
work and energy. To recover the Jarzynski relation we set
$\gamma=\beta$ and $\delta =0$ in Eq.~(\ref{13}) which then
assumes the form
\begin{equation}
\label{14} S (\beta; 0 ,t)=\sum_{m} \textit{G}_{m} (\beta,t)
\end{equation}
where
\begin{equation}
\label{15} \textit{G}_{m}(\beta,t)=\sum_{n}G_{mn}(\beta,t)
\frac{exp(-\beta\epsilon_{n}(0))}{Z_0}
\end{equation}
Let us consider the (unnormalized) distribution of states $m$
\begin{equation}
\label{16}
\textit{G}_{m}(\beta,t)=\frac{exp(-\beta\epsilon_{m}(t))}{Z_0}
\end{equation}
It can be easily verified that Eq.~(\ref{16}) is a solution to
Eq.~(\ref{12}); this follows from the detailed balance of $R$ that
guarantees that the first term in the r.h.s. of Eq.~(\ref{12})
vanishes when acting on Eq.~(\ref{16}). Substituting
Eq.~(\ref{16}) in Eq.~(\ref{14}) immediately gives the Jarzynski
relation (Eq.~(\ref{9a})). It will be interesting to extend these
results to systems described by more elaborate kinetic schemes
such as fractional kinetics~\cite{10} rather than the ordinary
master equation.

The validity of the Jarzynski relation for a system coupled to a
bath at constant temperature has been established by
Jarzynski~\cite{3} and Crooks~\cite{6}. The present derivation and
the definition of work closely resemble Jarzynski's master
equation approach. However, the path integral picture of
continuous segments with discrete jumps provides a new insight for
quantum systems and helps establish an interesting analogy with
the theory of pure dephasing of quantum coherence, as will be
shown next.

Eq.~(\ref{12}) closely resembles the stochastic Liouville equation
of Kubo~\cite{7,8,12}, which has been widely used in the theory of
NMR and optical lineshapes as well as in single molecule
spectroscopy~\cite{14}. Computing the work generating function is
then formally equivalent to computing the dephasing of a two level
system ($g$ and $e$) coupled to a bath. In that problem we
consider equilibrium fluctuations of a quantum coherence $\sigma$
between the two levels which are coupled to a classical bath
represented by a phase space $\Gamma$. $\sigma$ satisfies the
stochastic Liouville equation~\cite{7}
\begin{equation}
\label{new a} \frac{d \sigma(\Gamma, t)}{dt}= \sum_{\Gamma'} W
(\Gamma, \Gamma') \sigma (\Gamma', t)-i U (\Gamma) \sigma (\Gamma,
t).
\end{equation}
Here $W$ is a Markovian rate matrix and $U (\Gamma)$ is the
fluctuating energy gap between the two levels. $\sigma$ serves as
a generating function for the absorption lineshape which is given
by the Fourier transform of its zero'th moment $J (t) = \int
\sigma ( \Gamma, t)d \Gamma$. The normalization of $\sigma$ is
thus the physically interesting quantity whose Fourier transform
gives the spectral lineshape.

Eq.~(\ref{12}) and Eq.~(\ref{new a}) describe very different
physical phenomena: The former corresponds to an externally driven
nonequilibrium system whereas the latter represents a non driven
system at equilibrium. However, the two equations have a close
formal connection. To better see the analogy we recast the
solution of Eq.~(\ref{new a}) as
\begin{equation}
\label{new b} \sigma (\Gamma, t)= \langle exp\,\, [-i \int_0^t
d\tau U (\tau)]\rangle
\end{equation}
where the averaging is over a subensemble of trajectories which
assume the same value $\Gamma$ at time $t$. For comparison we
write the solution of Eq.~(\ref{12}) as
\begin{equation}
\label{20} G_m (\gamma, t) = \langle exp \,\, \left[-\gamma
\int_0^t d \tau
\partial\epsilon/\partial\tau\right]\rangle.
\end{equation}
The average here is over a subensemble of trajectories (Fig. 1)
which start at thermal equilibrium at $t=0$ and end up in state
$m$ at time $t$. We reiterate that $\partial\epsilon/\partial\tau$
only corresponds to the continuous part of the trajectory; the
jumps should be excluded.

In both equations we compute a phase space distribution whose time
dependent normalization gives the generating function for the
quantity of interest (either work or absorption spectrum). The
first term in the r.h.s. of Eq.~(\ref{12})) or (\ref{new a})
represents a regular dynamics which conserves the normalization,
whereas the second term is responsible for varying the
normalization, which is the physically interesting property. In
Eq.~(\ref{12}) or Eq.~(\ref{20}) this comes from the flow of
adiabatic state populations in and out of the system, originating
from the time dependence of adiabatic energies. In the dephasing
problem (Eq.~(\ref{new a}) or (\ref{new b})), the coherence
$\sigma$ acquires a random phase $(U)$ which is different for
different trajectories and it thus constitutes path function. This
translates to loss of magnitude (dephasing and decoherence) upon
ensemble averaging. The work in the nonadiabatic dynamics is
related to an integral over the continuous part of the energy
trajectory in the same way that the phase in the lineshape problem
is obtained by an integral over the fluctuating two level
frequency. This analogy suggests that it should be possible to
carry out nonequilibrium lineshape measurements by e.g. switching
on an electric field for a solute in a polar solvent and observing
the spectroscopic analogues of the Jarzynski relations by looking,
for example, at the time dependent Stokes Shift.

\vspace{1cm}

\textbf{Acknowledgement:}

The support of the National Science Foundation grant no.
CHE-0132571 is gratefully acknowledged. I wish to thank the
referee for most useful comments that helped clarify the
generality of these results.


\newpage

\begin{figure}[htbp]
\includegraphics[width=11.5cm, angle=-90]{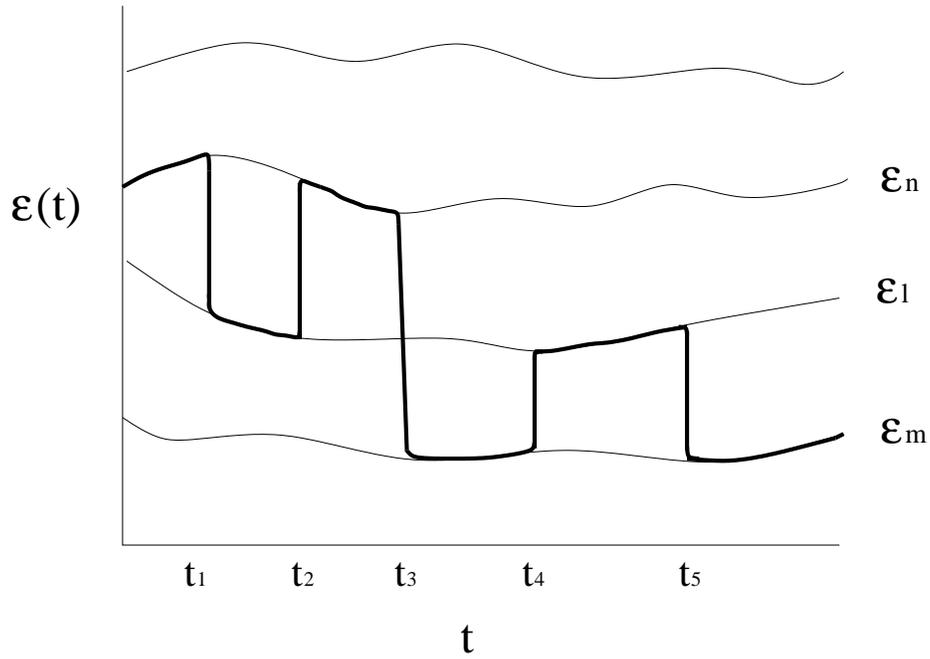}
\caption{A ``classical trajectory'' in population space of
adiabatic states. Shown is a trajectory which starts at state
$|\varphi_{n}(0)\rangle$ and ends in state
$|\varphi_{m}(t)\rangle$ and undergoes $j=5$ jumps at time
$t_{1}.....t_{5}$. The solution of the master equation is given by
a path integral over all possible trajectories (sum over all
possible values of $j=0,1,2.....$ and integration over $t_{j}$) }
\end{figure}
\clearpage


\newpage
\begin{thebibliography}{99}

\bibitem{1} C. Jarzynski, \textit{Phys.Rev. Let.}, \textbf{78}, 2690
(1997).

\bibitem{3} C. Jarzynski, \textit{Phys. Rev. E}, \textbf{56}, 5018 (1997).

\bibitem{2} G. Hummer, A. Szabo, \textit{PNAS}, \textbf{98}, 3658 (2001).

\bibitem{4} J. Liphardt, S. Dumont, S. B. Smith, I. Tinoco, C.
Bustamante, \textit{Science}, \textbf{296}, 1832 (2002).

\bibitem{6} G. E. Crooks, \textit{Phys. Rev.E}, \textbf{61}, 2361
(2000); G. E. Crooks \textit{J. Stat. Phys.} \textbf{90}, 1481
(1998).

\bibitem{5} D. A. Egolf, \textit{Science}, \textbf{296}, 1813 (2002).

\bibitem{9}A. Shapere and F. Wilczek, Ed. \emph{``Geometric Phases in Physics''},
World Scientific, Singapore (1989).

\bibitem{13} N. G. Van Kampen, \emph{``Stochastic Processes in Physics
and Chemistry''}, North Holland, New York (1984).

\bibitem{10} I. M. Sokolov, J. Klafter and A. Blumen, \textit{Phyics
Today}, November 2002 p.48; R. Metzler and J. Klafter,
\textit{Phys. Rep.}, \textbf{339}, 1-77, (2000).

\bibitem{7} R. Kubo, \textit{Adv. Chem. Phys.} \textbf{15}, 101
(1969); R. Kubo,  \textit{J. Math. Phys.} \textbf{4}, 174 (1963).

\bibitem{8} S. Mukamel, ``Principles of Nonlinear Optical
Spectroscopy'', Oxford University Press, New York (1995);
Paperback edition (1999).

\bibitem{12} Y. J. Yan, S. Mukamel, \textit{J Chem. Phys.},\textbf{ 88}, 5735
(1988).

\bibitem{14} ``Spectroscopy of Single Molecules in Physics,
Chemistry and Life Sciences'' \textit{Chem. Phys.} \textbf{247}, 1
(1999).
\end{thebibliography}
\end{document}